# Cavity nonlinear optics with layered materials

Taylor Fryett[1], Alan Zhan[2], Arka Majumdar[1,2,*]
[1] Electrical Engineering, University of Washington, Seattle, WA-98195
[2] Physics Department, University of Washington, Seattle, WA-98195
* corresponding author: arka@uw.edu

**Abstract**

Unprecedented material compatibility and ease of integration, in addition to the unique and diverse optoelectronic properties of layered materials have generated significant interest in their utilization in nanophotonic devices. While initial nanophotonic experiments primarily focused on light-sources, modulators, and detectors, recently researchers have demonstrated nonlinear optical devices using layered materials. In this paper, we review the current state of cavity-enhanced nonlinear optics with layered materials. Along with conventional nonlinear optics related to harmonic generation, we report on emerging directions of nonlinear optics, where the layered materials can potentially play a significant role.

**1. Introduction**

Layered materials have recently emerged as a promising class of materials for optoelectronics [7]. These materials exist with different band gaps leading to a large range of conductivities, as well as the ability to emit and detect light at different wavelength ranges [7]. For example, graphene has no band gap and behaves like a metal; semiconducting transition metal dichalcogenides (TMDCs, such as $WS_2$, $MoS_2$, $MoSe_2$ and $WSe_2$) and black phosphorous (BP) have band gaps on the order of $1 - 2.5\ eV$, and with its very wide band gap ($\sim 6\ eV$) hexagonal boron nitride (h-BN) is an excellent insulator. Going beyond usual conductivity, exotic properties, like superconductivity ($TaS_2$, $NbSe_2$) [10, 11], ferroelectricity ($CuInP_2S_6$) [13] and ferromagnetism ($Cr_2Ge_2Te_6$, $CrI_3$) [14-16] have been recently reported in these atomically thin materials. Figs. 1a,b show the schematic of the atomic structure for a typical TMDC and graphene. While the bulk of the research in these materials has focused on material characterization, optical spectroscopy and electrical transport measurements, integration of these materials with nanophotonic and nanoelectronic devices is gaining attention. Such integration efforts are primarily motivated by the ease of transferring these extremely thin materials on any substrate without needing explicit lattice matching. Unlike many other quantum confined structures such as, III-V quantum wells or self-assembled quantum dots, these materials stick to the substrate via van-der Waals force, and do not require expensive molecular beam epitaxy processes, or wafer bonding. This provides a unique opportunity to develop hybrid photonic platforms where the fabrication processes of the active and passive devices can be completely decoupled. Large-scale, passive photonic integrated circuits will be fabricated using existing foundry systems [17], and subsequently layered materials will be transferred on the pre-fabricated photonic chip to build the active devices, including modulators, detectors, light-sources (lasers or light emitting diodes) and nonlinear optical switches.

Over the last decade, the growth and extensive characterization of these materials have revealed their potential for various electronic and optoelectronic applications. Graphene, inarguably the most studied layered material [18], has found applications in building transistors [19], radio-frequency devices [20], and electronic sensors [21]. In the optoelectronics community, the large electro-optic tunability [22, 23] and strong photo-absorption [24] of graphene generated great interest. However, the extreme thinness of the material, the source of the unusual properties and unprecedented material compatibility of these layered materials also limits the effective light-matter interaction for optoelectronic device applications. To solve this problem, optical device engineers have used various nanophotonic devices integrated with graphene to enhance the lightmatter interaction. For example, using nanophotonic waveguides integrated with graphene, the effective light-matter interaction can be increased to build an electro-optic modulator [25, 26], or a photo-detector with high efficiency [27-29]. Further light-matter enhancement and reduction in the size of the device is possible using optical resonators, including photonic crystals [30, 31], ring resonators [26] and plasmonic resonators [32-34]. Optical resonators or cavities enhance the light-matter interaction by confining the light both temporally and spatially. For a cavity, the extent of the temporal and spatial confinement are quantitatively given by the quality factor ($Q$) and mode volume ($V$), respectively [35]. This trend of integrating graphene with nanophotonic devices continued with TMDCs [36-38]. Cavity integrated TMDCs have been used to demonstrate optically pumped lasing [39-41], cavity enhanced electroluminescence [42] and strongly coupled exciton-polaritons [3, 43]. In all these devices, the enhancement due to the cavity depends on the Purcell factor $F_p \sim Q/V$ [35, 44].

The success of optical resonators to enhance the light-matter interaction in layered materials motivated researchers to study cavity nonlinear optics using these materials. Due to their bosonic and charge-less nature, the interaction strength between photons is very weak and nonlinear optical effects appear only at very high power. This remains an outstanding challenge for optics to be used for any computing applications. However, the required power can be lowered using optical resonators. For a cavity enhanced second-order $\chi^{(2)}$ (third-order $\chi^{(3)}$) nonlinear switch, the threshold power is proportional to $V/Q^3$ ($V/Q^2$) [45]. The ability of nano-cavities to spatially confine light of wavelength $\lambda$ to an extremely small mode volume ($V \sim (\lambda/n)^3$ where $n$ is the refractive index of the cavity material), and store light for few nano-seconds ($\tau_{ph} = Q\lambda/2\pi c$) provides an opportunity to realize optical nonlinearity at the few photon level. Thus, cavities can significantly enhance the intrinsic optical nonlinearity of the layered materials. In this paper, we review the recent progress in the field of cavity nonlinear optics with layered materials. We note that the nonlinearity of layered materials is comparable to that achievable of exiting material systems, including quantum confined structures and bulk materials. However, the ability to integrate these materials with any material systems provides an excellent opportunity to build hybrid integrated nonlinear photonic systems. Apart from the cavity-enhanced second- and third-order nonlinear optics, the layered materials exhibit saturable absorption and nonlinear polariton-polariton interaction. Finally, the layered materials host defect centers, which are essentially two-level systems, and will provide a route to obtain single photon nonlinearity. Along with reviewing the existing works, we will lay out the outstanding challenges and several future directions of the cavity nonlinear optics with these emerging layered material systems.

Table 1: Reported nonlinear optical coefficients in layered materials: SI unit for $\chi^{(3)}$-coefficients is $m^2/V^2$ and for $\chi^{(2)}$-coefficients is $m/V$.

| Material [Ref] | Type of NLO ($\chi^{(3)}, \chi^{(2)}$) | Approx. $\lambda (nm)$ | Value (SI unit) |
|---|---|---|---|
| Graphene [46] | $\chi^{(3)}$ | 800 | $1.4 \times 10^{-15}$ |
| Graphene [47] | $\chi^{(3)}$ | 1550 | $3.25 \times 10^{-19}$ |
| BP [48] | $\chi^{(3)}$ | 1550 | $1.4 \times 10^{-19}$ |
| $MoS_2$ [49] | $\chi^{(3)}$ | 1550 | $2.9 \times 10^{-19}$ |
| GaSe [50] | $\chi^{(3)}$ | 1550 | $1.7 \times 10^{-16}$ |
| $ReS_2$ [51] | $\chi^{(3)}$ | 1550 | $5.3 \times 10^{-18}$ |
| $WSe_2$ [52] | $\chi^{(2)}$ | 1550 | $60 \times 10^{-12}$ |
| $WS_2$ [53] | $\chi^{(2)}$ | 800 | $10 \times 10^{-9}$ |
| $MoS_2$ [54] | $\chi^{(2)}$ | 800 | $100 \times 10^{-9}$ |
| $WSe_2$ [55] | $\chi^{(2)}$ | 800 | $10 \times 10^{-9}$ |
| h-BN [56] | $\chi^{(2)}$ | 800 | $1 \times 10^{-11}$ |
| GaSe [57] | $\chi^{(2)}$ | 1550 | $700 \times 10^{-12}$ |

## 2. Nonlinear processes in layered materials

In recent years, several experiments with stand-alone layered materials measured their nonlinear optical coefficients. Table 1 summarizes the measured $\chi^{(2)}$ and $\chi^{(3)}$ coefficients for different layered materials. The large $\chi^{(2)}$ values near ~800 $nm$ originate from the energetic proximity of the excitons in TMDCs, and thus generally come with large losses [2, 53-55, 58]. These experiments also clearly showed that the second order nonlinearity in TMDCs is the strongest at the single layer limit. With an even number of layers this nonlinearity disappears, as the material becomes centro-symmetric. Thus, the nonlinear signal works as a convenient microscopy tool for probing the number of layers in the TMDC crystal. Additionally, a six-fold polarization pattern is found in the second harmonic signal reflecting the hexagonal crystal structure of the TMDCs (Figs. 1c, d). Thus, nonlinear spectroscopy of the layered materials has become an indispensable tool to determine the crystalline orientation, thickness uniformity, layer stacking, and single-crystal domain size of atomically thin films [54, 56, 59, 60]. In fact, this technique is critical for creating the heterostructures of TMDC layers, as the relative crystal orientation determines the heterostructure's electronic band structure [61]. Researchers have also demonstrated that the centro-symmetry of bilayer $MoS_2$ can be broken with a static electric field, enabling second harmonic generation [62]. Note that, this electric field induced second harmonic generation (EFISH) can potentially create opportunities for electro-optic switching applications. Another way to break the symmetry will be via straining the bilayers. Straining monolayers and bilayers of $MoS_2$, researchers have already demonstrated band gap engineering and photoluminescence control [63-65]. However, no report exists on controlling the nonlinear optical properties yet. In addition to experimental measurements, several research groups have reported atomistic simulation studies to estimate the nonlinear coefficients for different layered materials [66-68], and have also analyzed the effect of substrates [69]. This work models the nonlinearity of layered materials as zero-thickness interfaces, and showed that without careful theoretical modelling the extracted value of the nonlinear coefficient can be largely exaggerated.

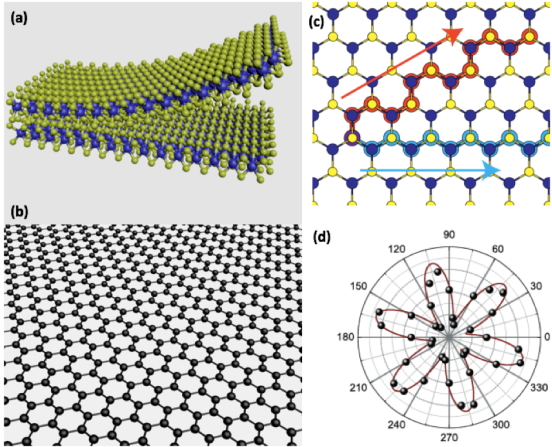

Figure 1: Schematic of (a) layered TMDC material and (b) graphene. (c) in TMDC, the crystal axes determine the nonlinear optical properties. The two directions shown are the zig-zag (show in blue) and arm-chair (shown in red) directions. (d) Due to the six-fold symmetry, a six-fold pattern is observed in polarization resolved second harmonic generation experiment. Thus, based on the second harmonic signal, the crystal structures of TMDCs can be confirmed. Fig. 1d is reprinted from the Ref. [2].

Most studies with TMDCs have focused on pumping near the exciton resonance (~800 nm) and creating second harmonic above the band gap at ~400 nm. A more interesting regime from an application point of view is to pump near 1550 nm and creating second harmonic signal at 775 nm. This wavelength range is particularly important for interfacing with telecommunication technologies. Using this process, we can realize optical bistability for a telecommunication wavelength laser [70]. In this wavelength range, the absorptive loss at the fundamental frequency is also minimal. Researchers have demonstrated second harmonic generation under pump near 1550 nm in stand-alone, monolayer $WSe_2$ [52] and in GaSe [57]. In $WSe_2$, researchers also demonstrated electric field controlled second-harmonic generation showing the nonlinear optical effect far from the excitonic frequency is strongest at the twice the exciton frequency. Thus, by changing the exciton frequency, one can change the effective nonlinearity. The reported $\chi^{(2)}$ values for $WSe_2$ are around 60pm/V at 1550 nm and for GaSe, the $\chi^{(2)}$ value is estimated to be at least one order of magnitude larger (Table 1).

In addition to second order nonlinear processes, third order nonlinear processes have also been studied using layered materials. Being centrosymmetric, graphene does not possess intrinsic second order nonlinearity, and hence graphene is primarily studied for third-order nonlinear processes. Several research papers explore third harmonic generation [46], cascaded third harmonic generation [71], and four wave mixing [72]. Layer-dependent third harmonic generation is also recently reported in black phosphorous [48]. In TMDCs, researchers have demonstrated intensity dependent refractive index by exploiting the third-order nonlinearity [73].

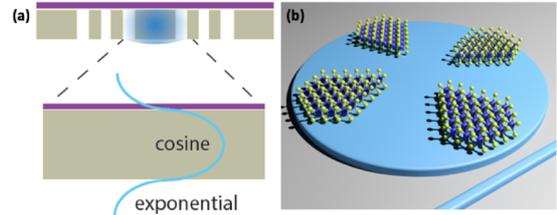

Figure 2: (a) The theoretical model to estimate the effective nonlinearity for a layered material placed on top of the cavity assumes the cavity mode follows a cosine function inside the slab, and decays exponentially outside the slab. (b) Schematic of a disk with patterned layered material on top: by intelligently patterning the material on top, the modes profiles are not significantly altered, but the nonlinear overlap can be significantly improved.

We note that, the measured values of the nonlinear coefficients in infrared wavelengths (~1500 nm) are comparable to the $\chi^{(3)}$ of bulk silicon ($\sim 10^{-19}\ m^2/V^2$) or the $\chi^{(2)}$ of bulk III-V materials, such as GaAs or GaP ($\sim 100 \times 10^{-12}\ m/V$), as a research article has pointed out in the past [74]. Thus, the utility of nonlinear layered materials does not necessarily originate from a large nonlinearity per se, but rather from their tunable optical nonlinearity, their ease of integration onto arbitrary substrates, ease of patterning, and the possibility of creating a hybrid platform. For example, CMOS-compatible materials, such as silicon and silicon nitride lack second-order optical nonlinearity. By integrating layered materials on top of them, we can realize second-order nonlinearity in a CMOS compatible way. Thus, in our opinion, the true benefit of layered materials lies in creating hybrid, active photonic systems.

## 3. Cavity enhanced second harmonic generation

While there is significant research progress in the nonlinear optics of layered materials, the extreme thinness of the material poses a serious problem for practical applications. This problem can be ameliorated by using an optical resonator to enhance the effective nonlinearity. While the effective nonlinearity in a layered material clad cavity is proportional to the material nonlinearity, the material interacts only with a small portion of the field on the surface. Hence it is important to estimate the effective nonlinearity in the presence of an optical resonator.

### 3.1 Modeling of the nonlinear cavity

In a recent work, Majumdar et. al. analyzed the effective nonlinearity of a layered material clad nano-cavity for second-order nonlinear optics [75]. They assume that the optical cavity is formed in a dielectric slab with the field profile varying sinusoidally inside the slab, and decaying exponentially outside (Fig. 2a). While such model is a simplification, and neglects the actual mode profiles of the cavity, it allows the authors to obtain a closed form analytical expression, which provides an intuitive understanding of the effective nonlinearity. They find that the nonlinear interaction strength $g_{nl}$ is given by [75]:

$$\hbar g_{nl} \approx \left(\frac{\hbar\omega_o}{2\varepsilon_o}\right)^{3/2} \frac{2\chi^{(2)} dS^3}{3\varepsilon^{3/2}\sqrt{\pi\sigma_x\sigma_y}}$$

where $\hbar$ is the reduced Planck's constant; $\omega_o$ is the fundamental angular frequency, $\chi^{(2)}$ is the value of the second-order nonlinear co-efficient, $d$ is the thickness of the layered materials; $\varepsilon$ is the dielectric constant of the layered material; $S$ signifies the ratio of the field strength at the surface and center of the slab; $\sigma_x$ and $\sigma_y$ are the confinement length of the optical field inside the resonator along $x$ and $y$ directions. Note that, a stronger localization within the nonlinear material will increase the nonlinear interaction strength. A cavity helps to confine light and thus reduces both $\sigma_x$ and $\sigma_y$. They also find that the interaction strength depends on the product of the nonlinear coefficient and the material thickness, as one would expect. Thus, the small thickness of the layered materials indeed limits the interaction strength. However, the analytical expression also shows that by enhancing the field on the surface, i.e., by increasing the value of $S$, the effective nonlinearity can be significantly enhanced. This criterion necessitates rethinking the design of the dielectric optical cavities, where the field is generally minimized on the surface to reduce the loss. In a more recent paper, the authors have also shown that layered materials provide a novel way to satisfy phase-matching condition in second-order nonlinear optics as elaborated below.

In deriving the analytical expression of the nonlinear interaction strength, it was previously assumed that the phase matching condition is fully satisfied. In second-order nonlinear optics the phase-matching condition is equivalent to satisfying the momentum conservation of the interacting waves [76]. In a nano-cavity, satisfying this condition amounts to having a good modal overlap between the fundamental and the second harmonic modes. Moreover, for high efficiency, the cavity needs to support modes at both frequencies. Satisfying both of these requirements is difficult, and to date, the proposed designs require stringent fabrication tolerance, and often involve significant computational resources [77]. A single monolayer does not significantly change the confined mode-profiles in the cavities as the interaction is evanescent in nature, and due to the extreme thinness of the material, only slightly perturbs the cavity mode. Thus, 2D materials provide an excellent opportunity to decouple the cavity design and nonlinear material design. Starting with any cavity, and intelligently patterning 2D materials on top of it, the phase-matching condition can be satisfied. A schematic of a nano-resonator with patterned 2D material on top is shown in Fig. 2b. Our theoretical analysis shows that such patterning can maintain a good overlap integral, and thus retain significant effective nonlinearity, even when the modes are largely mismatched [78].

The model described above primarily works for a dielectric optical resonator, and currently most of the reported results on cavity nonlinear optics with layered materials involve dielectric resonators. However, plasmonic resonators already are being used to enhance the linear properties of the layered materials [79], including light emission [40, 80], electro-optic modulation [33] and photo-detection [81]. Recently, evidence of polariton formation has also been observed in plasmonic cavities [82]. Plasmonic resonators can also be engineered to have multiple resonances [80], and thus can be ideal for second and third harmonic generation. While the quality factors of plasmonic resonances are significantly lower compared to dielectric resonators due to metallic losses, they provide an extremely small mode volume to enhance the light-matter interaction. Additionally, plasmonic effects can be exploited to confine light to lie in a single plane, where the layered material can be placed. Thus, plasmonic resonators are expected to play an important role in enhancing nonlinear optical effects in layered materials. Another important research direction could be to develop hybrid plasmonic-dielectric resonators to realize both high Q as is characteristic of dielectric cavities, with the low V that is characteristic of plasmonic cavities.

### 3.2 Second harmonic generation with single mode cavity

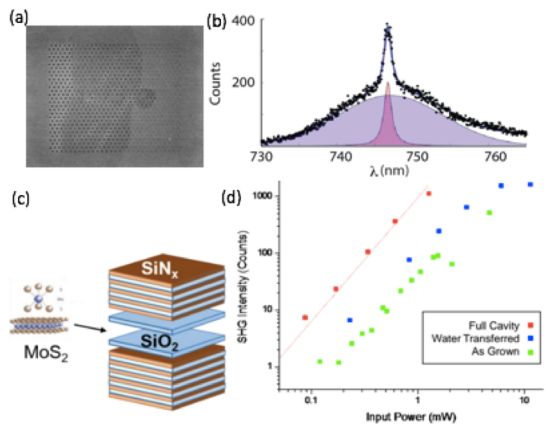

Figure 3: Cavity enhanced second-order nonlinear optics with layered materials: (a) SEM of a WSe$_2$-clad silicon photonic crystal cavity. (b) Measured spectrum of the second harmonic signal shows the Gaussian background with a Lorentzian peak signifying the cavity resonance. (c) Schematic of a DBR cavities with MoS$_2$ placed in between two mirrors. The DBR is made of alternating layers of silicon dioxide and silicon nitride. (d) The measured intensity of the second harmonic generated light shows a clear enhancement due to the cavity. Figures (a)-(b) are reprinted from the Ref. [4]; Figures (c)-(d) are reprinted from the Ref. [8].

Current experimental efforts on second harmonic generation with layered materials are primarily focused on a single cavity mode coupled to exfoliated layered materials. Using a tungsten diselenide (WSe$_2$) clad silicon photonic crystal cavity Fryett et. al. showed enhanced second harmonic generation [4]. In this work they used a pulsed laser operating within the telecommunication band (~1550 nm). Fig. 3a shows the scanning electron micrograph (SEM) of a WSe$_2$ clad silicon photonic crystal cavity. The cavity modes are near 1490 nm. When the cavity is resonantly excited using a pulsed laser, they observed a strong second harmonic signal around 745 nm (Fig. 3b). In the second harmonic signal, they observed a Gaussian background which comes from the doubling of the laser wavelengths, which are off-resonant from the cavity. In this background, they observed a Lorentzian peak at exactly the half-wavelength of the cavity resonance, signifying the cavity enhanced second-harmonic generation. The reported enhancement is only ~100, primarily because of the lack of a cavity mode at the second-harmonic frequency and low Q-factor of the cavity ($Q \sim 100$). Moreover, silicon absorbs a significant amount of second harmonic signal. A promising solution will be to use make a doubly resonant cavity out of wide bandgap materials, such as silicon nitride [38] or silicon dioxide [83]. Researchers also recently demonstrated continuous wave second harmonic generation using GaSe coupled with silicon photonic crystal cavity with a pump laser at ~1550 nm [84]. The required optical power in this work is only microwatts, primarily due to the high Q- factor and small mode-volume of the photonic crystal cavities.

In another experiment, Day et. al. demonstrated second harmonic generation using MoS$_2$ integrated inside a distributed Bragg reflector (DBR) cavity [8] (Figs. 3c, d). In this experiment, the DBR is formed by using alternating layers of silicon nitride and silicon dioxide to minimize the absorption of light. Here the second harmonic generation is observed under pulsed excitation at 800 nm. The reported enhancement is ~10. The lower enhancement factor can be attributed to low Q-factor of ~20, and the large mode-volume of the DBR cavity. An open hemiconfocal cavity geometry [85] might be more suitable to reduce the mode volume and thus further enhance the nonlinearity.

### 3.3 Second harmonic generation with double mode cavity

The enhancement factor can be significantly increased by using cavity modes both at the fundamental and the second harmonic frequency. Such mode engineering is theoretically difficult at the nano-scale, but can be easily realized in a DBR-based Fabry-Perot cavity. However, inevitable fabrication errors prevent the cavity modes from appearing exactly at the desired resonance wavelengths. Yi et. al., solved this problem by creating a mechanically tunable Fabry-Perot cavity [12], where the bottom mirror is a DBR, and the top mirror is a capacitively actuated silver mirror (Fig. 4a). The capacitive tuning enables changing the cavity length, and thus cavity resonances. Via mechanical tuning, it is possible to bring both the fundamental and the second harmonic modes to the desired frequencies. In this experiment, they used MoS$_2$ as the nonlinear material, and despite their low cavity Q-factor, the reported enhancement is ~2000 (Figs. 4b,c). This experiment used a pulsed excitation near 930 nm. Further improvement is possible by improving the cavity Q-factor, and reducing the mode-volume, which is often very large in Fabry-Perot cavities.

Phase-matched second-harmonic generation was also recently reported in a silicon photonic waveguide using MoSe$_2$ [86]. By engineering the waveguide cross-section, the effective mode-indices of the fundamental and second-harmonic modes are matched which ensures the phase-matching of the light at the fundamental and second harmonic frequencies. Similar methods have been used for designing phase-matched ring resonators for cavity nonlinear optics using aluminum nitride [87]. Such a phase-matched ring integrated with layered materials can enhance the efficiency of the second-order nonlinear processes.

All the existing experiments so far have reported significantly lower SHG efficiency, compared to the theoretical predictions. For a given power, the efficiency can be improved by increasing the quality factor and reducing the mode volume. Another limiting factor might be the transfer of the layered materials, which generally involves organic polymers, and might introduce excess loss coming from the polymer residues. Having the resonances at both the fundamental and the second harmonic frequencies will also increase the efficiency significantly. Along with innovating cavity structures, new layered materials with stronger nonlinear optical properties, such as multiferroics [88] can also improve the overall efficiency of second harmonic generation.

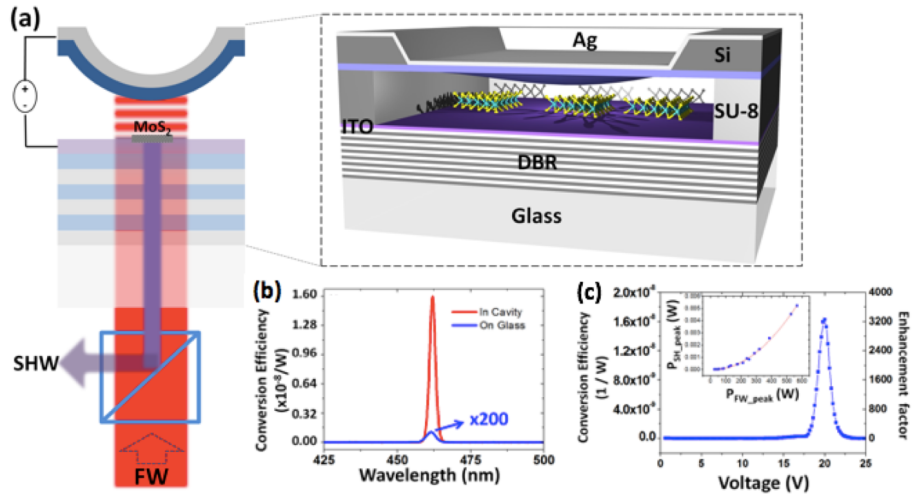

Figure 4: Second harmonic generation in double-mode cavity: (a) Schematic of the mechanically tunable Fabry-Perot cavity with embedded MoS$_2$. (b) The frequency conversion efficiency is increased due to the presence of the cavity. (c) By applying the voltage, the cavity modes can be tuned, and a large enhancement can be achieved when the fundamental resonance frequency and the twice of the second harmonic frequency become equal. Figures are reprinted with permission from the Ref. [12]. Copyright (2016) American Chemical Society.

### 4. Third order nonlinear processes

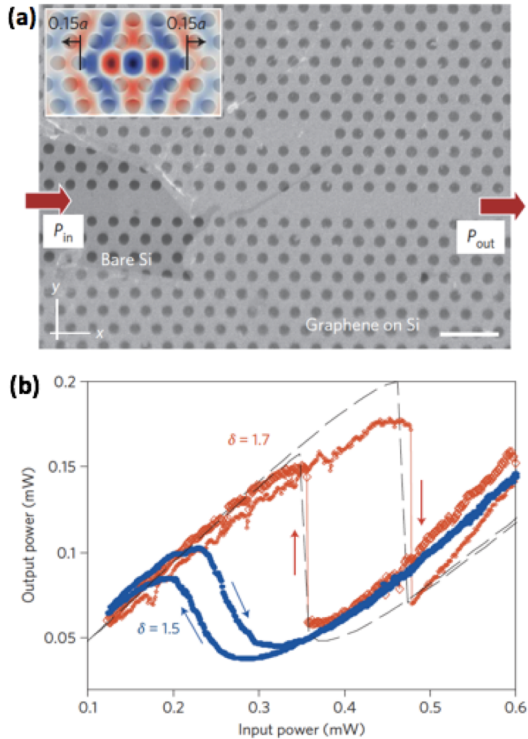

Figure 5: Optical bistability with graphene-clad silicon photonic crystal cavity: (a) SEM of the graphene-clad photonic crystal cavity; (b) Output power as a function of the input power clearly shows the signature of optical bistability. The figures are reprinted from the Ref. [1].

While most of the recent works on cavity nonlinear optics using layered materials primarily focused on second-order nonlinear optics, the first demonstration of cavity nonlinear optics was with third-order nonlinearity of graphene. Gu et. al. demonstrated optical bistability and regenerative oscillation using a graphene-clad silicon photonic crystal cavity [1]. Fig. 5a shows the fabricated photonic crystal cavity, which is coupled to a waveguide. Graphene is transferred onto the cavity, and under ~1562 nm CW excitation optical bistability is observed (Fig. 5b). The threshold power where the bistable behavior appears in graphene-silicon resonator is lower than what is observed with a bare silicon photonic crystal cavity.

In the authors' opinion, however, the third order nonlinearity with 2D materials is not as attractive as second-order nonlinear optics from an application point of view. As most materials already possess strong third-order nonlinearity, and the measured values of nonlinearity for graphene and BP are almost the same order of magnitude as silicon, the application area of third-order nonlinear optics with layered material is debatable [74]. However, for many applications, requiring high power operation, silicon is not ideal due to high two-photon absorption when operating at the telecommunication wavelength. In those applications, silicon nitride and silicon dioxide are gaining popularity due to their wide band gaps [89]. The magnitude of the third-order nonlinear coefficient in layered materials is higher compared to that of the nitride ($\chi^{(3)} \sim 1 \times 10^{-20}\ m^2/V^2$) [90] or oxide ($\chi^{(3)} \sim 2.5 \times 10^{-22}\ m^2/V^2$) [76], and will play an important role in realizing low-power nonlinear optics.

### 5. Saturable Absorption

Another promising nonlinear optical effect in layered materials is their saturable absorption. Due to the extreme thinness of layered materials, their photo-absorption can be easily saturated with low optical power. A common application of saturable absorption is the integration into optical cavities to generate pulsed laser sources, both in fiber and free space systems. Many recent papers on pulsed lasers have utilized layered 2D materials as the saturable absorption material of choice [9, 91, 92]. Most of these works used graphene as the saturable absorber. One particularly appealing aspect of graphene is its broadband absorption due to its lack of a band gap, which allows mode-locking over a large wavelength range, including in the mid-infrared wavelengths. Fig. 6a shows an experimental setup with a graphene saturable absorption mirror [5]. In this work, stable mode-locked laser pulses as short as 729 fs were obtained with a repetition rate of 98.7 MHz and an average output power of 60.2 mW at $\sim 2\ \mu m$ (Fig. 6b). In addition to the free-space setup, saturable absorption in graphene was used in a fiber laser to demonstrate mode locking (Fig. 6c) [9]. In this work, they created a passively mode-locked erbium-doped fiber laser working at 1559 nm, with 460 fs pulse duration (Fig. 6d). Recently, several works reported similar mode-locked laser systems with a variety of atomically thin materials including black phosphorous [93], MoS$_2$ [94] and WS$_2$ [95]. Additionally, researchers have used black phosphorous, WS$_2$ and MoS$_2$ solutions as saturable absorbers to construct passively Q-switched Nd:YVO$_4$ lasers, with pulse durations of few nanoseconds [96].

## 6. Nonlinear exciton-polaritons

So far, most of the nonlinear optical effects we discussed are observed with light sources off-resonant from the exciton. Very near the exciton resonances, a large nonlinearity can be realized using the polaritonic system formed by strong coupling between the cavity-confined photons, and the excitons in the layered materials. Recent works reported observation of strongly coupled exciton-polaritons using layered materials, including MoS$_2$ and bilayer MoSe$_2$ [3, 43]. Fig. 7a shows the schematic of the open DBR cavity used to demonstrate strong coupling between the cavity mode and the exciton [3]. The open cavity architecture allows tuning the cavity resonance by mechanically displacing one of the cavity mirrors. When the cavity mode is tuned across the excitonic resonance, anti-crossing between the cavity mode and the exciton is observed, signifying strong coupling and polariton formation. Fig. 7b shows the upper and lower polaritons measured in photoluminescence.

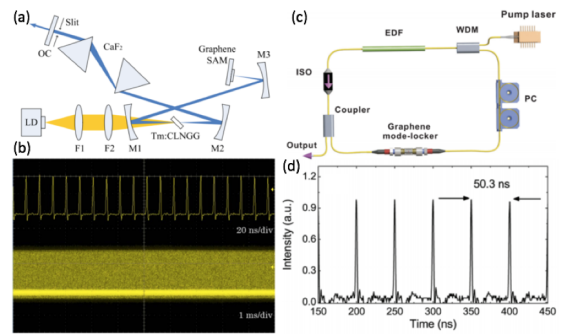

Figure 6: (a) Experimental setup of the mode-locked laser based on the graphene saturable absorption mirror; (b) CW mode-locked pulse trains in nanosecond and millisecond time scales; (c) Graphene mode-locked fiber laser where the mode-locker assembly contains a graphene flake. ISO: isolator; WDM: wavelength division multiplexer; PC: polarization controller; EDF: erbium-doped fiber. (d) Auto-correlation trace of output pulses showing the pulse repetition rate of 19MHz. Figures (a), (b) are reprinted from the Ref. [5]; Figures (c), (d) are reprinted with permission from the Ref. [9]. Copyright (2010) American Chemical Society.

Exciton-polaritons using quantum wells have been previously used to extensively study "quantum fluids of light" [97]. At low excitation power, the exciton-polariton system behaves linearly as the exciton-exciton interaction is weak. Bose-Einstein condensation (BEC) of exciton-polaritons has also been observed by several groups [98, 99]. The polaritons have much smaller effective mass compared to atoms due to their photonic components, and thus the BEC can be realized at a much higher temperature. Moreover, the extremely large binding energy ($\sim 0.2 - 0.8\ eV$) in TMDC excitons can potentially allow creation of polaritons and condensates at room temperature. Such condensation has not yet been observed in layered materials, but is theoretically predicted [100]. Theory also predicts observation of topological polaritons [101] and exciton-mediated superconductivity using the TMDC exciton-polaritons [102]. With higher excitation power, however, one cannot create an arbitrarily number of excitons [99]. Thus, the fermionic nature of the constituent electron and hole in the exciton-polariton becomes more prominent in high density exciton-polariton, and strong polariton-polariton repulsion can be observed. When the exciton density reaches the Mott density, the exciton-polariton system can potentially exhibit BEC-BCS crossover [99]. Thus, layered material based exciton-polaritons, and their condensates can potentially provide a strongly nonlinear material platform.

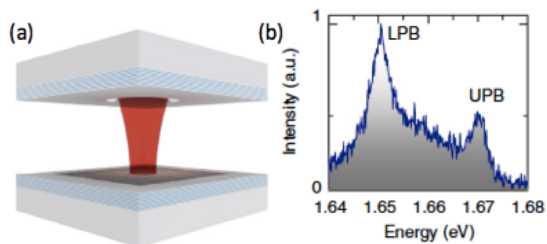

Figure 7: Exciton-polaritons in layered materials: (a) schematic of the open DBR cavity with layered material inside the cavity. The top mirror of the cavity can be mechanically displaced to tune the cavity resonance. (b) Photoluminescence spectrum of the strongly coupled exciton-polariton system clearly shows the upper and lower polaritons. The figures are reprinted from the Ref. [3].

Additionally, the extreme thinness of the layered materials allows for straightforward patterning by etching. Thus, one can easily pattern them to create small islands of layered materials, which behave like quantum dots (Fig. 8a). The size and position of these islands can be easily controlled by lithography [103]. Performance of such quantum dot like structures has been theoretically analyzed [104]. In TMDCs with excitonic Bohr radius of only 1 nm, theoretical analysis showed that when the radius of the patterned material reaches few 10's of nm, single photon nonlinear optics can be observed [105]. Specifically, polaritonic blockade can be realized in these quantum dot like structures. The ability to deterministically position these structures via lithography, can potentially create an array of interacting single photon nonlinear systems (Fig. 8b). Researchers have already designed and fabricated array of linear cavities [106, 107]. The addition of the single photon nonlinearity in each of these cavities will create a testbed to study non-equilibrium quantum many body physics with correlated photons [108-110].

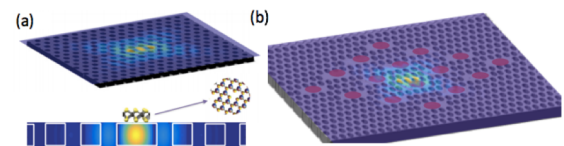

Figure 8: Patterned quantum dots in layered materials: (a) By patterning layered 2D materials, we can create quantum dot like structures, where the exciton-exciton interaction can be significantly enhanced. (b) These patterned quantum dots can be placed in an array and couple with an array of interacting optical resonators. Thus, a network of nonlinear nodes can be created.

## 7. Single photon nonlinear optics

Several theoretical proposals explored the feasibility of reaching single photon nonlinear optics using the second and third-order optical nonlinear materials, coupled to optical resonators [111, 112]. As layered materials can be integrated with high-Q silicon or silicon nitride cavities, it is indeed possible to reach single photon nonlinear optics using the off-resonant nonlinearity. However, the quality factor needed to reach the single photon nonlinear optical regime is generally very high $\sim 10^5 - 10^6$, and it is not yet clear whether such a high Q can be reached with polymer-based transfer process of layered materials. The creation of the quantum dot like structures by lithography are promising, but the effect of etching on the non-radiative excitonic decay rate is not clear.

A more straightforward route would be to exploit quantum emitters in layered materials. Recently several research groups have reported single quantum emitters in layered materials, particularly in TMDCs and in h-BN [6, 113-115]. These single emitters originate from the localized defects in the crystals [116]. Fig. 9a shows the scanning PL data from a CVD-grown WSe$_2$ flake, where the bright defects can be clearly identified. When such a bright spot is spatially separated, the measured spectrum shows narrow lines, indicative of single defects (Fig. 9b). Via second-order autocorrelation measurements under both CW and pulsed excitation (Figs. 9c, d), a $g^{(2)}(0) < 0.5$ is measured, which unambiguously proves single photon emission from these defects. Via coupling two-level quantum emitters, such as quantum dots, with an optical cavity, strong nonlinear optical effects, such as single photon switching [117-119], and photon blockade [120-122] have been observed. Similar performance is expected from the defects in layered materials as well. To this end, Tran et al. have coupled the single defects in h-BN to plasmonic resonators [123]. The plasmonic pillars enhanced the overall brightness of the emitters by a factor of two. Recently, enhancement of the quantum emitters in WSe$_2$ is also observed using silver nanowires, where the emission is coupled to a surface plasmon polariton mode [124]. Finally, very recently scalable growth of defects in layered material has been demonstrated [125]. In this work, monolayers of WSe$_2$ and WS$_2$ are transferred in a templated vertical silica nano-pillars, and with high probability the defects are localized near the pillar. The wavelength of the defect emission also can be controlled by the pillar diameter. Such capability of deterministic positioning and wavelength selection can solve the long-standing problem of stochastic positioning and large inhomogeneous broadening of quantum dots, which have largely limited the scalable operation of quantum dots, and other quantum emitters. The defects in layered materials thus can provide a scalable platform of single photon nonlinear devices.

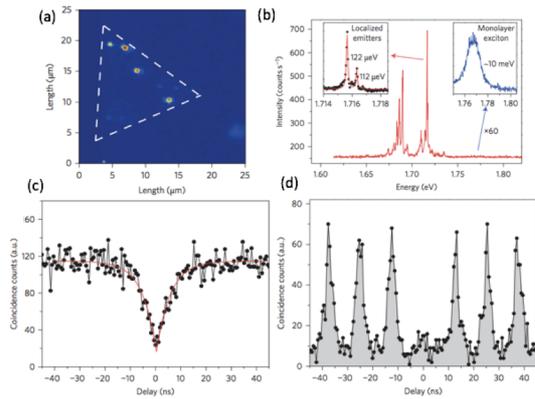

Figure 9: Single quantum emitters in layered materials: (a) photoluminescence intensity map of narrow emission lines over $25\ \mu m \times 25\ \mu m$ area. The dashed triangle indicates the position of the monolayer; (b) photoluminescence spectrum of localized emitters. The left inset is a high-resolution spectrum of a defect emission. The right inset is a zoom-in of the monolayer valley exciton emission; Observation of photon anti-bunching with (c) cw and (d) pulsed excitation. The dip at zero time-delay indicates presence of single photons. This figure is reprinted from the Ref. [6].

## 8. Outlook

The field of the cavity nonlinear optics with layered materials is in its infancy. The majority of researchers working with layered materials still primarily focus on the spectroscopy and material characterization, and the effort on integration with nanophotonic resonant structures is relatively recent. However, the unprecedented material compatibility, easy availability of the materials, and unique optoelectronic properties of the layered materials have very quickly generated strong interest in the nanophotonics community. Hence, we are hopeful that integration of layered materials with nanophotonic devices will enable more fundamentally new scientific studies, and novel low-power applications in the near future. As explained earlier, all the reported nonlinearities can be enhanced by improving the cavities (larger $Q$ and smaller $V$), and this will surely drive future research on cavity nonlinear optics with layered material systems. In this section, we elaborate some of the new and more speculative research directions involving cavity nonlinear optics with layered materials.

### 8.1  Phase transition layered materials

Recently phase transition in layered materials, such as $MoTe_2$ and $WTe_2$ has been predicted [126-128]. Bulk phase change materials, such as GeSbTe (GST) have recently generated a lot of interest in the field of integrated nanophotonics due to the large change in the refractive index associated with the structural change in GST [129-131]. Coupled with an optical resonator, such large change in refractive index will enable optical bistability, where the output optical power is a strongly nonlinear function of the input optical power. One major problem of GST, however, is large amount of loss in the visible and near-infrared frequency. The layered phase change materials have a wider band gap, and can provide a new way to realize a very large change in the refractive indices, while maintaining a low loss. This change coupled with optical cavities can provide a large nonlinear effect.

### 8.2  Heterostructure of layered materials

Recently several groups have reported fabrication and characterization of heterostructures of layered materials [132, 133]. One particularly interesting aspect of the heterostructure is the observation of long-lived inter-layer excitons, where the electron and holes are confined in two different layered materials and hence are physically located in two different spatial locations. This prevents the recombination of the electron and hole pair, and thus long-lived exciton states can be realized. The implication of such long-lived excitons for nonlinear optics is not clear, and can potentially constitute a new research field. For example, what are the implication for exciton-polaritons and exciton-polaritons BEC with long-lived excitons? Further, by separating different TMDC layers by h-BN, a multi-quantum well like structure can be realized, where we essentially increase the thickness of the nonlinear material. Already researchers have demonstrated stronger exciton-photon coupling in bi-layer $MoSe_2$ separated by h-BN [3]. With $N$ layers, we expect the coupling strength to increase by a factor of $\sqrt{N}$.

### 8.3  Valley exciton-polariton

TMDCs exhibit unique spin-valley physics, which primarily stems from the combination of two features. First, the band gap is at the $+K$ and $-K$ "Dirac valleys", not at the Brillouin zone center. This gives the electron states a "valley index" (or pseudospin) in addition to real spin [134, 135]. Optical selection rules are such that right hand circularly polarized light ($\sigma^+$) couples only to one valley and left hand circularly polarized light ($\sigma^-$) couples to the other, providing the first solid-state system in which dynamical control of valley pseudospin is possible [134, 136-143]. Second, the strong spin-orbit coupling locks the real spin at the band edges to the valley index [134]. Spontaneous transitions between these spin-states are unlikely due to the large mismatch between momentum vectors. Recently several research groups have observed valley physics in the exciton-polariton system [144-146]. Surprisingly, the results show that the hybridization with cavity photons makes it easier to observe the spin-valley physics at room temperature [144], and for materials, where the exciton does not ordinarily demonstrate strong spin-valley physics [146]. However, the spin-valley physics remained largely ignored in the cavity nonlinear optics with monolayer materials. In III-V quantum well systems, researchers demonstrated helicity dependent polariton-polariton interaction. Exploiting such effect, multistability in helicity is observed in exciton-polariton system [147, 148]. Similar effects are expected with TMDC exciton-polaritons as well. The multistable devices can enable multi-state optical logic.

### 8.4  Hyperbolic metamaterial with layered materials

In all the nonlinear systems we have considered in this article so far, the layered materials only provide the required nonlinearity and the cavity is formed by another otherwise linear material. However, it is possible to fabricate a whole photonic structure using layered materials. For example, using graphene and h-BN layered structures researchers predicted that a hyperbolic metamaterial can be created [149, 150]. Hyperbolic metamaterials are promising candidates for enhancing the nonlinear optical interaction [151]. Thus, it is possible to build a nonlinear nanophotonic system solely made out of layered materials. This will open a completely new field of research.

## 9. Acknowledgement

The authors acknowledge useful discussion with Prof. Vinod Menon, Prof. Xiaodong Xu and Prof. Feng Wang. This work is supported by the National Science Foundation under grant NSF-EFRI-1433496, and the Air Force Office of Scientific Research-Young Investigator Program under grant FA9550-15-1-0150.

## 10. References

[1]	GuT, PetroneN, J. F. McMillan, A. van der Zande, YuM, G. Q. Lo*, et al.*, "Regenerative oscillation and four-wave mixing in graphene optoelectronics," *Nat Photon,* vol. 6, pp. 554-559, 2012.

[2] L. M. Malard, T. V. Alencar, A. P. M. Barboza, K. F. Mak, and A. M. de Paula, "Observation of intense second harmonic generation from MoS2 atomic crystals," *Physical Review B,* vol. 87, p. 201401, 05/13/ 2013.
[3] S. Dufferwiel, S. Schwarz, F. Withers, A. A. P. Trichet, F. Li, M. Sich*, et al.*, "Exciton-polaritons in van der Waals heterostructures embedded in tunable microcavities," *Nature communications,* vol. 6, p. 8579, 2015.
[4] T. K. Fryett, K. L. Seyler, J. Zheng, C.-H. Liu, X. Xu, and A. Majumdar, "Silicon photonic crystal cavity enhanced second-harmonic generation from monolayer WSe2," *arXiv:1607.03548,* 2016.
[5] J. Ma, G. Q. Xie, P. Lv, W. L. Gao, P. Yuan, L. J. Qian*, et al.*, "Graphene mode-locked femtosecond laser at 2micron wavelength," *Optics Letters,* vol. 37, pp. 2085-2087, 2012/06/01 2012.
[6] Y.-M. He, ClarkGenevieve, R. SchaibleyJohn, Y. He, C. ChenMing, J. WeiYu*, et al.*, "Single quantum emitters in monolayer semiconductors," *Nat Nano,* vol. 10, pp. 497-502, 06//print 2015.
[7] F. Xia, H. Wang, D. Xiao, M. Dubey, and A. Ramasubramaniam, "Two-dimensional material nanophotonics," *Nat Photon,* vol. 8, pp. 899-907, 12//print 2014.
[8] J. K. Day, M.-H. Chung, Y.-H. Lee, and V. M. Menon, "Microcavity enhanced second harmonic generation in 2D MoS2," *Optical Materials Express,* vol. 6, pp. 2360-2365, 2016/07/01 2016.
[9] Z. Sun, T. Hasan, F. Torrisi, D. Popa, G. Privitera, F. Wang*, et al.*, "Graphene Mode-Locked Ultrafast Laser," *ACS Nano,* vol. 4, pp. 803-810, 2010/02/23 2010.
[10] X. Xi, Z. Wang, W. Zhao, J.-H. Park, K. T. Law, H. Berger*, et al.*, "Ising pairing in superconducting NbSe2 atomic layers," *Nat Phys,* vol. 12, pp. 139-143, 02//print 2016.
[11] E. Navarro-Moratalla, J. O. Island, S. Mañas-Valero, E. Pinilla-Cienfuegos, A. Castellanos-Gomez, J. Quereda*, et al.*, "Enhanced superconductivity in atomically thin TaS2," vol. 7, p. 11043, 03/17/online 2016.
[12] F. Yi, M. Ren, J. C. Reed, H. Zhu, J. Hou, C. H. Naylor*, et al.*, "Optomechanical Enhancement of Doubly Resonant 2D Optical Nonlinearity," *Nano Letters,* vol. 16, pp. 1631-1636, 2016/03/09 2016.
[13] F. Liu, L. You, K. L. Seyler, X. Li, P. Yu, J. Lin*, et al.*, "Room-temperature ferroelectricity in CuInP2S6 ultrathin flakes," vol. 7, p. 12357, 08/11/online 2016.
[14] P. Je-Geun, "Opportunities and challenges of 2D magnetic van der Waals materials: magnetic graphene?," *Journal of Physics: Condensed Matter,* vol. 28, p. 301001, 2016.
[15] C. Gong, L. Li, Z. Li, H. Ji, A. Stern, Y. Xia*, et al.*, "Discovery of intrinsic ferromagnetism in two-dimensional van der Waals crystals," *Nature,* vol. 546, pp. 265-269, 06/08/print 2017.
[16] B. Huang, G. Clark, E. Navarro-Moratalla, D. R. Klein, R. Cheng, K. L. Seyler*, et al.*, "Layer-dependent ferromagnetism in a van der Waals crystal down to the monolayer limit," *Nature,* vol. 546, pp. 270-273, 06/08/print 2017.
[17] M. Hochberg and T. Baehr-Jones, "Towards fabless silicon photonics," *Nat Photon,* vol. 4, pp. 492-494, 2010.
[18] A. K. Geim and K. S. Novoselov, "The rise of graphene," *Nat Mater,* vol. 6, pp. 183-191, 2007.
[19] F. Schwierz, "Graphene transistors," *Nat Nano,* vol. 5, pp. 487-496, 07//print 2010.
[20] H. Wang, A. Hsu, J. Wu, J. Kong, and T. Palacios, "Graphene-Based Ambipolar RF Mixers," *IEEE Electron Device Letters,* vol. 31, pp. 906-908, 2010.
[21] F. Schedin, A. K. Geim, S. V. Morozov, E. W. Hill, P. Blake, M. I. Katsnelson*, et al.*, "Detection of individual gas molecules adsorbed on graphene," *Nat Mater,* vol. 6, pp. 652-655, 09//print 2007.
[22] F. Wang, Y. Zhang, C. Tian, C. Girit, A. Zettl, M. Crommie*, et al.*, "Gate-Variable Optical Transitions in Graphene," *Science,* vol. 320, pp. 206-209, April 11, 2008 2008.
[23] Z. Fei, A. S. Rodin, G. O. Andreev, W. Bao, A. S. McLeod, M. Wagner*, et al.*, "Gate-tuning of graphene plasmons revealed by infrared nano-imaging," *ArXiv:1202.4993v2,* 2012.
[24] R. R. Nair, P. Blake, A. N. Grigorenko, K. S. Novoselov, T. J. Booth, T. Stauber*, et al.*, "Fine Structure Constant Defines Visual Transparency of Graphene," *Science,* vol. 320, p. 1308, June 6, 2008 2008.
[25] M. Liu, X. Yin, E. Ulin-Avila, B. Geng, T. Zentgraf, L. Ju*, et al.*, "A graphene-based broadband optical modulator," *Nature,* vol. 474, pp. 64-67, 2011.
[26] C. T. Phare, Y.-H. Daniel Lee, J. Cardenas, and M. Lipson, "Graphene electro-optic modulator with 30 GHz bandwidth," *Nat Photon,* vol. 9, pp. 511-514, 08//print 2015.
[27] X. Gan, R.-J. Shiue, Y. Gao, I. Meric, T. F. Heinz, K. Shepard*, et al.*, "Chip-integrated ultrafast graphene photodetector with high responsivity," *Nat Photon,* vol. 7, pp. 883-887, 2013.
[28] T. Mueller, F. Xia, and P. Avouris, "Graphene photodetectors for high-speed optical communications," *Nat Photon,* vol. 4, pp. 297-301, 05//print 2010.
[29] F. Xia, T. Mueller, Y.-m. Lin, A. Valdes-Garcia, and P. Avouris, "Ultrafast graphene photodetector," *Nat Nano,* vol. 4, pp. 839-843, 2009.
[30] X. Gan, R.-J. Shiue, Y. Gao, K. F. Mak, X. Yao, L. Li*, et al.*, "High-Contrast Electrooptic Modulation of a Photonic Crystal Nanocavity by Electrical Gating of Graphene," *Nano Letters,* vol. 13, pp. 691-696, 2013/02/13 2013.
[31] A. Majumdar, J. Kim, J. Vuckovic, and F. Wang, "Electrical Control of Silicon Photonic Crystal Cavity by Graphene," *Nano Letters,* vol. 13, pp. 515-518, 2013/02/13 2013.
[32] W. Wei, J. Nong, Y. Zhu, L. Tang, G. Zhang, J. Yang*, et al.*, "Cavity-enhanced continuous graphene plasmonic resonator for infrared sensing," *Optics Communications,* vol. 395, pp. 147-153, 7/15/ 2017.
[33] J. Kim, H. Son, D. J. Cho, B. Geng, W. Regan, S. Shi*, et al.*, "Electrical Control of Optical Plasmon Resonance with Graphene," *Nano Letters,* vol. 12, pp. 5598-5602, 2012/11/14 2012.
[34] A. Majumdar, K. Jonghwan, J. Vuckovic, and W. Feng, "Graphene for Tunable Nanophotonic Resonators," *Selected Topics in Quantum Electronics, IEEE Journal of,* vol. 20, pp. 68-71, 2014.
[35] K. J. Vahala, "Optical microcavities," *Nature,* vol. 424, pp. 839-846, 2003.
[36] X. Gan, Y. Gao, K. Fai Mak, X. Yao, R.-J. Shiue, A. van der Zande*, et al.*, "Controlling the spontaneous emission rate of monolayer MoS2 in a photonic crystal nanocavity," *Applied Physics Letters,* vol. 103, p. 181119, 2013.
[37] S. Wu, S. Buckley, A. M. Jones, J. S. Ross, N. Ghimire, J., J. Yan*, et al.*, "Control of two-dimensional excitonic light emission via photonic crystal," *2D Materials,* vol. 1, p. 011001, 2014.
[38] G. Wei, T. Stanev, N. Stern, D. Czaplewski, and W. Jung, "Interfacing Monolayer MoS2 with Silicon-Nitride Integrated Photonics," in *Advanced Photonics 2015*, Boston, Massachusetts, 2015, p. IM4A.3.
[39] S. Wu, S. Buckley, J. R. Schaibley, L. Feng, J. Yan, D. G. Mandrus*, et al.*, "Monolayer semiconductor nanocavity lasers with ultralow thresholds," *Nature,* vol. 520, pp. 69-72, 2015.
[40] Y. Ye, Z. J. Wong, Xiufang Lu, X. Ni, H. Zhu, X. Chen*, et al.*, "Monolayer Excitonic Laser," *Nature Photonics,* vol. 9, pp. 733–737, 2015.
[41] O. Salehzadeh, M. Djavid, N. H. Tran, I. Shih, and Z. Mi, "Optically Pumped Two-Dimensional MoS2 Lasers Operating at Room-Temperature," *Nano Letters,* vol. 15, pp. 5302-5306, 2015/08/12 2015.
[42] C.-H. Liu, G. Clark, T. Fryett, S. Wu, J. Zheng, F. Hatami*, et al.*, "Nanocavity Integrated van der Waals Heterostructure Light-Emitting Tunneling Diode," *Nano Letters,* 2016/12/07 2016.

[43] X. Liu, T. Galfsky, Z. Sun, F. Xia, E.-c. Lin, Y.-H. Lee, *et al.*, "Strong light–matter coupling in two-dimensional atomic crystals," *Nat Photon,* vol. 9, pp. 30-34, 01//print 2015.
[44] K. Liu, S. Sun, A. Majumdar, and V. J. Sorger, "Fundamental Scaling Laws in Nanophotonics," *Scientific Reports,* vol. 6, p. 37419, 11/21/online 2016.
[45] R. Trivedi, U. K. Khankhoje, and A. Majumdar, "Cavity-Enhanced Second-Order Nonlinear Photonic Logic Circuits," *Physical Review Applied,* vol. 5, p. 054001, 05/02/ 2016.
[46] S.-Y. Hong, J. I. Dadap, N. Petrone, P.-C. Yeh, J. Hone, and R. M. Osgood, "Optical Third-Harmonic Generation in Graphene," *Physical Review X,* vol. 3, p. 021014, 06/10/ 2013.
[47] J. L. Cheng, N. Vermeulen, and J. E. Sipe, "Third order optical nonlinearity of graphene," *New Journal of Physics,* vol. 16, p. 053014, 2014.
[48] N. Youngblood, R. Peng, A. Nemilentsau, T. Low, and M. Li, "Layer-Tunable Third-Harmonic Generation in Multilayer Black Phosphorus," *ACS Photonics,* vol. 4, pp. 8-14, 2017/01/18 2017.
[49] R. I. Woodward, R. T. Murray, C. F. Phelan, R. E. P. d. Oliveira, T. H. Runcorn, E. J. R. Kelleher, *et al.*, "Characterization of the second- and third-order nonlinear optical susceptibilities of monolayer MoS 2 using multiphoton microscopy," *2D Materials,* vol. 4, p. 011006, 2017.
[50] L. Karvonen, A. Säynätjoki, S. Mehravar, R. D. Rodriguez, S. Hartmann, D. R. T. Zahn, *et al.*, "Investigation of Second- and Third-Harmonic Generation in Few-Layer Gallium Selenide by Multiphoton Microscopy," vol. 5, p. 10334, 05/19/online 2015.
[51] Q. Cui, R. A. Muniz, J. E. Sipe, and H. Zhao, "Strong and anisotropic third-harmonic generation in monolayer and multilayer ${\mathrm{ReS}}_{2}$," *Physical Review B,* vol. 95, p. 165406, 04/06/ 2017.
[52] K. L. Seyler, J. R. Schaibley, P. Gong, P. Rivera, A. M. Jones, S. Wu, *et al.*, "Electrical control of second-harmonic generation in a WSe2 monolayer transistor," *Nat Nano,* vol. 10, pp. 407-411, 05//print 2015.
[53] C. Janisch, Y. Wang, D. Ma, N. Mehta, A. L. Elías, N. Perea-Lopez, *et al.*, "Extraordinary Second Harmonic Generation in Tungsten Disulfide Monolayers," *Sci. Rep.,* vol. 4, 2014.
[54] N. Kumar, S. Najmaei, Q. Cui, F. Ceballos, P. M. Ajayan, J. Lou, *et al.*, "Second harmonic microscopy of monolayer MoS2," *Physical Review B,* vol. 87, p. 161403, 04/15/ 2013.
[55] J. Ribeiro-Soares, C. Janisch, Z. Liu, A. L. Elías, M. S. Dresselhaus, M. Terrones, *et al.*, "Second Harmonic Generation in WSe 2," *2D Materials,* vol. 2, p. 045015, 2015.
[56] Y. Li, Y. Rao, K. F. Mak, Y. You, S. Wang, C. R. Dean, *et al.*, "Probing Symmetry Properties of Few-Layer MoS2 and h-BN by Optical Second-Harmonic Generation," *Nano Letters,* vol. 13, pp. 3329-3333, 2013/07/10 2013.
[57] X. Zhou, J. Cheng, Y. Zhou, T. Cao, H. Hong, Z. Liao, *et al.*, "Strong Second-Harmonic Generation in Atomic Layered GaSe," *Journal of the American Chemical Society,* vol. 137, pp. 7994-7997, 2015/07/01 2015.
[58] H. Zeng, G.-B. Liu, J. Dai, Y. Yan, B. Zhu, R. He, *et al.*, "Optical signature of symmetry variations and spin-valley coupling in atomically thin tungsten dichalcogenides," *Sci. Rep.,* vol. 3, 2013.
[59] X. Yin, Z. Ye, D. A. Chenet, Y. Ye, K. O'Brien, J. C. Hone, *et al.*, "Edge Nonlinear Optics on a MoS2 Atomic Monolayer," *Science,* vol. 344, pp. 488-490, May 2, 2014 2014.
[60] W.-T. Hsu, Z.-A. Zhao, L.-J. Li, C.-H. Chen, M.-H. Chiu, P.-S. Chang, *et al.*, "Second Harmonic Generation from Artificially Stacked Transition Metal Dichalcogenide Twisted Bilayers," *ACS Nano,* vol. 8, pp. 2951-2958, 2014/03/25 2014.
[61] H. Heo, J. H. Sung, S. Cha, B.-G. Jang, J.-Y. Kim, G. Jin, *et al.*, "Interlayer orientation-dependent light absorption and emission in monolayer semiconductor stacks," vol. 6, p. 7372, 06/23/online 2015.
[62] J. Klein, J. Wierzbowski, A. Steinhoff, M. Florian, M. Rösner, F. Heimbach, *et al.*, "Electric-Field Switchable Second-Harmonic Generation in Bilayer MoS2 by Inversion Symmetry Breaking," *Nano Letters,* vol. 17, pp. 392-398, 2017/01/11 2017.
[63] H. J. Conley, B. Wang, J. I. Ziegler, R. F. Haglund, S. T. Pantelides, and K. I. Bolotin, "Bandgap Engineering of Strained Monolayer and Bilayer MoS2," *Nano Letters,* vol. 13, pp. 3626-3630, 2013/08/14 2013.
[64] A. Castellanos-Gomez, R. Roldán, E. Cappelluti, M. Buscema, F. Guinea, H. S. J. van der Zant, *et al.*, "Local Strain Engineering in Atomically Thin MoS2," *Nano Letters,* vol. 13, pp. 5361-5366, 2013/11/13 2013.
[65] C. R. Zhu, G. Wang, B. L. Liu, X. Marie, X. F. Qiao, X. Zhang, *et al.*, "Strain tuning of optical emission energy and polarization in monolayer and bilayer MoS${}_{2}$," *Physical Review B,* vol. 88, p. 121301, 09/09/ 2013.
[66] M. L. Trolle, G. Seifert, and T. G. Pedersen, "Theory of excitonic second-harmonic generation in monolayer MoS2," *Physical Review B,* vol. 89, p. 235410, 06/10/ 2014.
[67] M. Grüning and C. Attaccalite, "Second harmonic generation in $h$-BN and MoS${}_{2}$ monolayers: Role of electron-hole interaction," *Physical Review B,* vol. 89, p. 081102, 02/03/ 2014.
[68] C.-Y. Wang and G.-Y. Guo, "Nonlinear Optical Properties of Transition-Metal Dichalcogenide MX2 (M = Mo, W; X = S, Se) Monolayers and Trilayers from First-Principles Calculations," *The Journal of Physical Chemistry C,* vol. 119, pp. 13268-13276, 2015/06/11 2015.
[69] M. Merano, "Nonlinear optical response of a two-dimensional atomic crystal," *Optics Letters,* vol. 41, pp. 187-190, 2016/01/01 2016.
[70] T. Fryett, C. M. Dodson, and A. Majumdar, "Cavity Enhanced Nonlinear Optics for Few Photon Optical Bistability," *Optics Express,* vol. 23, pp. 16246-16255, 2015.
[71] D. A. Smirnova and A. S. Solntsev, "Cascaded third-harmonic generation in hybrid graphene-semiconductor waveguides," *Physical Review B,* vol. 92, p. 155410, 10/07/ 2015.
[72] E. Hendry, P. J. Hale, J. Moger, A. K. Savchenko, and S. A. Mikhailov, "Coherent Nonlinear Optical Response of Graphene," *Physical Review Letters,* vol. 105, p. 097401, 08/26/ 2010.
[73] G. Wang, S. Zhang, X. Zhang, L. Zhang, Y. Cheng, D. Fox, *et al.*, "Tunable nonlinear refractive index of two-dimensional MoS2, WS2, and MoSe2 nanosheet dispersions [Invited]," *Photonics Research,* vol. 3, pp. A51-A55, 2015/04/01 2015.
[74] J. B. Khurgin, "Graphene—A rather ordinary nonlinear optical material," *Applied Physics Letters,* vol. 104, p. 161116, 2014.
[75] A. Majumdar, C. M. Dodson, T. K. Fryett, A. Zhan, S. Buckley, and D. Gerace, "Hybrid 2D Material Nanophotonics: A Scalable Platform for Low-Power Nonlinear and Quantum Optics," *ACS Photonics,* vol. 2, pp. 1160-1166, 2015/08/19 2015.
[76] R. W. Boyd, *Nonlinear Optics,* Third ed.: Academic Press.
[77] Z. Lin, X. Liang, M. Lončar, S. G. Johnson, and A. W. Rodriguez, "Cavity-enhanced second-harmonic generation via nonlinear-overlap optimization," *Optica,* vol. 3, pp. 233-238, 2016/03/20 2016.
[78] T. Fryett, A. Zhan, and A. Majumdar, "Phase matched nonlinear optics via patterning layered materials," *arXiv:1706.09392,* 2017.
[79] X. Li, J. Zhu, and B. Wei, "Hybrid nanostructures of metal/two-dimensional nanomaterials for plasmon-enhanced applications," *Chemical Society Reviews,* vol. 45, pp. 3145-3187, 2016.
[80] G. M. Akselrod, T. Ming, C. Argyropoulos, T. B. Hoang, Y. Lin, X. Ling, *et al.*, "Leveraging Nanocavity Harmonics for Control of Optical Processes in 2D Semiconductors," *Nano Letters,* vol. 15, pp. 3578-3584, 2015/05/13 2015.
[81] C. Chen, N. Youngblood, R. Peng, D. Yoo, D. A. Mohr, T. W. Johnson, *et al.*, "Three-Dimensional Integration of Black Phosphorus Photodetector with Silicon Photonics and Nanoplasmonics," *Nano Letters,* vol. 17, pp. 985-991, 2017/02/08 2017.
[82] B. Lee, W. Liu, C. H. Naylor, J. Park, S. C. Malek, J. S. Berger, *et al.*, "Electrical Tuning of Exciton–Plasmon Polariton Coupling in Monolayer MoS2 Integrated with Plasmonic Nanoantenna Lattice," *Nano Letters,* 2017/06/14 2017.
[83] S. Hammer, H. M. Mangold, A. E. Nguyen, D. Martinez-Ta, S. Naghibi Alvillar, L. Bartels, *et al.*, "Scalable and Transfer-Free Fabrication of MoS2/SiO2 Hybrid Nanophotonic Cavity Arrays with Quality Factors Exceeding 4000," *Scientific Reports,* vol. 7, p. 7251, 2017/08/03 2017.

[84] X. Gan, C. Zhao, S. Hu, T. Wang, Y. Song, J. Li, *et al.*, "Microwatts continuous-wave pumped second harmonic generation in few- and mono-layer GaSe," *arXiv:1706.07923,* 2017.
[85] P. R. Dolan, G. M. Hughes, F. Grazioso, B. R. Patton, and J. M. Smith, "Femtoliter tunable optical cavity arrays," *Optics Letters,* vol. 35, pp. 3556-3558, 2010/11/01 2010.
[86] H. Chen, V. Corboliou, A. S. Solntsev, D.-Y. Choi, M. A. Vincenti, D. Ceglia, *et al.*, "Enhanced second-harmonic generation from two-dimensional MoSe2 by waveguide integration," in *Conference on Lasers and Electro-Optics*, San Jose, California, 2017, p. FM2F.4.
[87] X. Guo, C.-L. Zou, H. Jung, and H. X. Tang, "On-Chip Strong Coupling and Efficient Frequency Conversion between Telecom and Visible Optical Modes," *Physical Review Letters,* vol. 117, p. 123902, 09/16/ 2016.
[88] H. Wang and X. Qian, "Giant Optical Second Harmonic Generation in Two-Dimensional Multiferroics," *Nano Letters,* vol. 17, pp. 5027-5034, 2017/08/09 2017.
[89] W. H. P. Pernice, M. Li, D. F. G. Gallagher, and H. X. Tang, "Silicon nitride membrane photonics," *Journal of Optics A: Pure and Applied Optics,* vol. 11, p. 114017, 2009.
[90] K. Ikeda, R. E. Saperstein, N. Alic, and Y. Fainman, "Thermal and Kerr nonlinear properties of plasma-deposited silicon nitride/silicon dioxide waveguides," *Optics Express,* vol. 16, pp. 12987-12994, 2008/08/18 2008.
[91] J. Ma, G. Q. Xie, P. Lv, W. L. Gao, P. Yuan, L. J. Qian, *et al.*, "Graphene mode-locked femtosecond laser at 2 micron wavelength," *Optics Letters,* vol. 37, pp. 2085-2087, 2012/06/01 2012.
[92] Q. Bao, H. Zhang, Y. Wang, Z. Ni, Y. Yan, Z. X. Shen, *et al.*, "Atomic-Layer Graphene as a Saturable Absorber for Ultrafast Pulsed Lasers," *Advanced Functional Materials,* vol. 19, pp. 3077-3083, 2009.
[93] Y. Chen, G. Jiang, S. Chen, Z. Guo, X. Yu, C. Zhao, *et al.*, "Mechanically exfoliated black phosphorus as a new saturable absorber for both Q-switching and Mode-locking laser operation," *Optics Express,* vol. 23, pp. 12823-12833, 2015/05/18 2015.
[94] E. J. Aiub, D. Steinberg, E. A. Thoroh de Souza, and L. A. M. Saito, "200-fs mode-locked Erbium-doped fiber laser by using mechanically exfoliated MoS2 saturable absorber onto D-shaped optical fiber," *Optics Express,* vol. 25, pp. 10546-10552, 2017/05/01 2017.
[95] D. Mao, Y. Wang, C. Ma, L. Han, B. Jiang, X. Gan, *et al.*, "WS2 mode-locked ultrafast fiber laser," vol. 5, p. 7965, 01/22/online 2015.
[96] H. Liu, Z. Sun, X. Wang, Y. Wang, and G. Cheng, "Several nanosecond Nd:YVO4 lasers Q-switched by two dimensional materials: tungsten disulfide, molybdenum disulfide, and black phosphorous," *Optics Express,* vol. 25, pp. 6244-6252, 2017/03/20 2017.
[97] I. Carusotto and C. Ciuti, "Quantum fluids of light," *Reviews of Modern Physics,* vol. 85, pp. 299-366, 2013.
[98] T. Byrnes, N. Y. Kim, and Y. Yamamoto, "Exciton-polariton condensates," *Nat Phys,* vol. 10, pp. 803-813, 11//print 2014.
[99] H. Deng, H. Haug, and Y. Yamamoto, "Exciton-polariton Bose-Einstein condensation," *Reviews of Modern Physics,* vol. 82, pp. 1489-1537, 2010.
[100] J.-H. Jiang and S. John, "Photonic Architectures for Equilibrium High-Temperature Bose-Einstein Condensation in Dichalcogenide Monolayers," *Sci. Rep.,* vol. 4, 12/11/online 2014.
[101] T. Karzig, C.-E. Bardyn, N. H. Lindner, and G. Refael, "Topological Polaritons," *Physical Review X,* vol. 5, p. 031001, 07/01/ 2015.
[102] O. Cotleţ, S. Zeytinoğlu, M. Sigrist, E. Demler, and A. Imamoğlu, "Superconductivity and other collective phenomena in a hybrid Bose-Fermi mixture formed by a polariton condensate and an electron system in two dimensions," *Physical Review B,* vol. 93, p. 054510, 02/09/ 2016.
[103] G. Wei, D. A. Czaplewski, E. J. Lenferink, T. K. Stanev, I. W. Jung, and N. P. Stern, "Valley Polarization in Size-Tunable Monolayer Semiconductor Quantum Dots," *arXiv:1510.09135*.
[104] A. Verger, C. Ciuti, and I. Carusotto, "Polariton quantum blockade in a photonic dot," *Physical Review B,* vol. 73, p. 193306, 05/12/ 2006.
[105] J.-H. Jiang, H. Wang, A. Zhan, Y. Xu, H. Chen, and A. Majumdar, "Quantum Many-Body Simulation using Cavity Coupled Monolayer Excitons," *arXiv:1603.02394,* 2016.
[106] A. Majumdar, A. Rundquist, M. Bajcsy, V. D. Dasika, S. R. Bank, and J. Vučković, "Design and analysis of photonic crystal coupled cavity arrays for quantum simulation," *Physical Review B,* vol. 86, p. 195312, 2012.
[107] H. Altug, D. Englund, and J. Vuckovic, "Ultrafast photonic crystal nanocavity laser," *Nat Phys,* vol. 2, pp. 484-488, 2006.
[108] J. Eisert, M. Friesdorf, and C. Gogolin, "Quantum many-body systems out of equilibrium," *Nat Phys,* vol. 11, pp. 124-130, 02//print 2015.
[109] N. Changsuk and G. A. Dimitris, "Quantum simulations and many-body physics with light," *Reports on Progress in Physics,* vol. 80, p. 016401, 2017.
[110] J. H. Michael, "Quantum simulation with interacting photons," *Journal of Optics,* vol. 18, p. 104005, 2016.
[111] S. Ferretti and D. Gerace, "Single-photon nonlinear optics with Kerr-type nanostructured materials," *Physical Review B,* vol. 85, p. 033303, 2012.
[112] A. Majumdar and D. Gerace, "Single-photon blockade in doubly resonant nanocavities with second-order nonlinearity," *Physical Review B,* vol. 87, p. 235319, 2013.
[113] KoperskiM, NogajewskiK, AroraA, CherkezV, MalletP, J. Y. Veuillen, *et al.*, "Single photon emitters in exfoliated WSe2 structures," *Nat Nano,* vol. 10, pp. 503-506, 06//print 2015.
[114] A. Srivastava, M. Sidler, A. V. Allain, D. S. Lembke, A. Kis, and ImamoğluA, "Optically active quantum dots in monolayer WSe2," *Nat Nano,* vol. 10, pp. 491-496, 06//print 2015.
[115] T. T. Tran, K. Bray, M. J. Ford, M. Toth, and I. Aharonovich, "Quantum emission from hexagonal boron nitride monolayers," *Nat Nano,* vol. 11, pp. 37-41, 01//print 2016.
[116] L. Zhong, R. C. Bruno, K. Ethan, L. Ruitao, R. Rahul, T. Humberto, *et al.*, "Defect engineering of two-dimensional transition metal dichalcogenides," *2D Materials,* vol. 3, p. 022002, 2016.
[117] D. Englund, A. Majumdar, M. Bajcsy, A. Faraon, P. Petroff, and J. Vučković, "Ultrafast Photon-Photon Interaction in a Strongly Coupled Quantum Dot-Cavity System," *Physical Review Letters,* vol. 108, p. 093604, 2012.
[118] R. Bose, D. Sridharan, H. Kim, G. S. Solomon, and E. Waks, "Low-Photon-Number Optical Switching with a Single Quantum Dot Coupled to a Photonic Crystal Cavity," *Physical Review Letters,* vol. 108, p. 227402, 2012.
[119] T. Volz, A. Reinhard, M. Winger, A. Badolato, K. J. Hennessy, E. L. Hu, *et al.*, "Ultrafast all-optical switching by single photons," *Nat Photon,* vol. 6, pp. 605-609, 2012.
[120] A. Faraon, I. Fushman, D. Englund, N. Stoltz, P. Petroff, and J. Vuckovic, "Coherent generation of non-classical light on a chip via photon-induced tunnelling and blockade," *Nat Phys,* vol. 4, pp. 859-863, 2008.
[121] A. Majumdar, M. Bajcsy, and J. Vučković, "Probing the ladder of dressed states and nonclassical light generation in quantum-dot–cavity QED," *Physical Review A,* vol. 85, p. 041801, 2012.
[122] A. Reinhard, T. Volz, M. Winger, A. Badolato, K. J. Hennessy, E. L. Hu, *et al.*, "Strongly correlated photons on a chip," *Nat Photon,* vol. 6, pp. 93-96, 2012.
[123] T. T. Tran, D. Wang, Z.-Q. Xu, A. Yang, M. Toth, T. W. Odom, *et al.*, "Deterministic Coupling of Quantum Emitters in 2D Materials to Plasmonic Nanocavity Arrays," *Nano Letters,* vol. 17, pp. 2634-2639, 2017/04/12 2017.
[124] T. Cai, S. Dutta, S. Aghaeimeibodi, Z. Yang, S. Nah, J. T. Fourkas, *et al.*, "Coupling emission from single localized defects in 2D semiconductor to surface plasmon polaritons," *arXiv:1706.01532,* 2017.
[125] C. Palacios-Berraquero, D. M. Kara, A. R. P. Montblanch, M. Barbone, P. Latawiec, D. Yoon, *et al.*, "Large-scale quantum-emitter arrays in atomically thin semiconductors," vol. 8, p. 15093, 05/22/online 2017.

[126]	Y. Li, K.-A. N. Duerloo, K. Wauson, and E. J. Reed, "Structural semiconductor-to-semimetal phase transition in two-dimensional materials induced by electrostatic gating," vol. 7, p. 10671, 02/12/online 2016.
[127]	K.-A. N. Duerloo, Y. Li, and E. J. Reed, "Structural phase transitions in two-dimensional Mo- and W-dichalcogenide monolayers," vol. 5, p. 4214, 07/01/online 2014.
[128]	K.-A. N. Duerloo and E. J. Reed, "Structural Phase Transitions by Design in Monolayer Alloys," *ACS Nano,* vol. 10, pp. 289-297, 2016/01/26 2016.
[129]	C. Ríos, M. Stegmaier, P. Hosseini, D. Wang, T. Scherer, C. D. Wright*, et al.*, "Integrated all-photonic non-volatile multi-level memory," *Nat Photon,* vol. 9, pp. 725-732, 11//print 2015.
[130]	H. Liang, R. Soref, J. Mu, A. Majumdar, X. Li, and W.-P. Huang, "Simulations of Silicon-on-Insulator Channel-Waveguide Electrooptical 2x2 Switches and 1x1 Modulators Using a GeSeTe Self-Holding Layer," *Journal of Lightwave Technology,* vol. 33, pp. 1805-1813, 2015/05/01 2015.
[131]	P. Hosseini, C. D. Wright, and H. Bhaskaran, "An optoelectronic framework enabled by low-dimensional phase-change films," *Nature,* vol. 511, pp. 206-211, 07/10/print 2014.
[132]	A. Geim and I. Grigorieva, "Van der Waals heterostructures," *Nature,* vol. 499, pp. 419-425, 2013.
[133]	P. Rivera, J. R. Schaibley, A. M. Jones, J. S. Ross, S. Wu, G. Aivazian*, et al.*, "Observation of long-lived interlayer excitons in monolayer MoSe2–WSe2 heterostructures," *Nat Commun,* vol. 6, 02/24/online 2015.
[134]	D. Xiao, G.-B. Liu, W. Feng, X. Xu, and W. Yao, "Coupled Spin and Valley Physics in Monolayers of MoS2 and Other Group-VI Dichalcogenides," *Physical Review Letters,* vol. 108, p. 196802, 05/07/ 2012.
[135]	X. Xu, W. Yao, D. Xiao, and T. F. Heinz, "Spin and pseudospins in layered transition metal dichalcogenides," *Nat Phys,* vol. 10, pp. 343-350, 05//print 2014.
[136]	T. Cao, G. Wang, W. Han, H. Ye, C. Zhu, J. Shi*, et al.*, "Valley-selective circular dichroism of monolayer molybdenum disulphide," *Nature communications,* vol. 3, 2012 2012.
[137]	Hualing Zeng, Junfeng Dai, Wang Yao, Di Xiao, and a. X. Cui, "Valley polarization in MoS2 monolayers by optical pumping," *Nature Nanotechnology, doi:10.1038/nnano.2012.95,* 2012.
[138]	A. M. Jones, H. Yu, N. J. Ghimire, S. Wu, G. Aivazian, J. S. Ross*, et al.*, "Optical generation of excitonic valley coherence in monolayer WSe$_2$," *Nature Nanotechnology,* vol. 8, pp. 634-638, 2013.
[139]	Kin Fai Mak, Keliang He, Jie Shan, and T. F. Heinz, "Control of valley polarization in monolayer MoS2 by optical helicity," *Nature Nanotechnology, doi:10.1038/nnano.2012.96,* 2012.
[140]	Y. Li, J. Ludwig, T. Low, A. Chernikov, X. Cui, G. Arefe*, et al.*, "Valley Splitting and Polarization by the Zeeman Effect in Monolayer MoSe2," *arXiv:1409.8538* 2014.
[141]	A. Srivastava, M. Sidler, A. V. Allain, D. S. Lembke, A. Kis, and A. Imamoglu, "Valley Zeeman Effect in Elementary Optical Excitations of a Monolayer WSe2," *arXiv:1407.2624* 2014.
[142]	H. Zeng, J. Dai, W. Yao, D. Xiao, and X. Cui, "Valley polarization in MoS2 monolayers by optical pumping," *Nat Nano,* vol. 7, pp. 490-493, 08//print 2012.
[143]	D. Bajoni, P. Senellart, E. Wertz, I. Sagnes, A. Miard, A. Lemaître*, et al.*, "Polariton Laser Using Single Micropillar GaAs-GaAlAs Semiconductor Cavities," *Physical Review Letters,* vol. 100, p. 047401, 01/28/ 2008.
[144]	Y.-J. Chen, J. D. Cain, T. K. Stanev, V. P. Dravid, and N. P. Stern, "Valley-polarized exciton–polaritons in a monolayer semiconductor," *Nat Photon,* vol. 11, pp. 431-435, 07//print 2017.
[145]	Z. Sun, J. Gu, A. Ghazaryan, Z. Shotan, C. R. Considine, M. Dollar*, et al.*, "Optical control of room-temperature valley polaritons," *Nat Photon,* vol. 11, pp. 491-496, 08//print 2017.
[146]	DufferwielS, T. P. Lyons, D. D. Solnyshkov, A. A. P. Trichet, WithersF, SchwarzS*, et al.*, "Valley-addressable polaritons in atomically thin semiconductors," *Nat Photon,* vol. 11, pp. 497-501, 08//print 2017.
[147]	T. K. Paraïso, M. Wouters, Y. Léger, F. Morier-Genoud, and B. Deveaud-Plédran, "Multistability of a coherent spin ensemble in a semiconductor microcavity," *Nat Mater,* vol. 9, pp. 655-660, 08//print 2010.
[148]	N. A. Gippius, I. A. Shelykh, D. D. Solnyshkov, S. S. Gavrilov, Y. G. Rubo, A. V. Kavokin*, et al.*, "Polarization Multistability of Cavity Polaritons," *Physical Review Letters,* vol. 98, p. 236401, 06/04/ 2007.
[149]	DaiS, MaQ, M. K. Liu, AndersenT, FeiZ, M. D. Goldflam*, et al.*, "Graphene on hexagonal boron nitride as a tunable hyperbolic metamaterial," *Nat Nano,* vol. 10, pp. 682-686, 08//print 2015.
[150]	A. Kumar, T. Low, K. H. Fung, P. Avouris, and N. X. Fang, "Tunable Light–Matter Interaction and the Role of Hyperbolicity in Graphene–hBN System," *Nano Letters,* vol. 15, pp. 3172-3180, 2015/05/13 2015.
[151]	A. Poddubny, I. Iorsh, P. Belov, and Y. Kivshar, "Hyperbolic metamaterials," *Nat Photon,* vol. 7, pp. 948-957, 12//print 2013.